\documentclass[preprint,showpacs,preprintnumbers,amsmath,amssymb,nofootinbib]{revtex4}
\usepackage{bbm}
\usepackage{amsfonts}
\usepackage{booktabs}
\usepackage{mathrsfs}
\usepackage{epsfig}
\usepackage{graphicx}
\usepackage{dcolumn}
\usepackage{bm}
\usepackage{amsmath}
\usepackage{slashed}

\let\jnfont=\rm
\def\NPB#1,{{\jnfont Nucl.\ Phys.\ B }{\bf #1},}
\def\PLB#1,{{\jnfont Phys.\ Lett.\ B }{\bf #1},}
\def\EPJC#1,{{\jnfont Eur.\ Phys.\ Jour.\ C }{\bf #1},}
\def\PRD#1,{{\jnfont Phys.\ Rev.\ D }{\bf #1},}
\def\PRL#1,{{\jnfont Phys.\ Rev.\ Lett.\ }{\bf #1},}
\def\MPLA#1,{{\jnfont Mod.\ Phys.\ Lett.\ A }{\bf #1},}
\def\JPG#1,{{\jnfont J.\ Phys.\ G}{\bf #1},}
\def\CTP#1,{{\jnfont Commun.\ Theor.\ Phys.\ }{\bf #1},}
\def\ZPC#1,{{\jnfont Z.\ Phys.\ C }{\bf #1},}
\def\JHEP#1,{{\jnfont JHEP \ }{\bf #1},}
\def\Rv{\not{\hbox{\kern-1pt $R$}}}
\def\p{\not{\hbox{\kern-3pt $p$}}}

\begin{document}

\title{Probing topcolor-assisted technicolor
       from top charge asymmetry and triple-top production
       at the LHC}

\author{Chengcheng Han$^{1,2}$, Ning Liu$^1$, Lei Wu$^2$, Jin Min Yang$^2$
        \\~ \vspace*{-0.3cm} }
\affiliation{ $^1$ Physics Department, Henan Normal University,
     Xinxiang 453007, China\\
$^2$ State Key Laboratory of Theoretical Physics, Institute of Theoretical Physics,
     Academia Sinica, Beijing 100190, China
     \vspace*{1.5cm}}

\begin{abstract}
In a topcolor-assisted technicolor model (TC2) with large FCNC top
quark couplings, we study its correlated contributions to the top
quark forward-backward asymmetry ($A_{FB}$) at the Tevatron, the top
charge asymmetry ($A_{C}$) and the triple-top production at the LHC.
Under current constraints on the top quark from the LHC and Tevatron
(such as the total and differential production rates), we scan the
parameter space of such a TC2 model. We find that in the allowed
parameter space the TC2 model can explain the Tevatron measured
$A_{FB}$ at $2\sigma$ level, but meanwhile significantly enhance
$A_{C}$ at the LHC. Such enhanced $A_{C}$, albeit currently allowed
by the LHC measurement at $2\sigma$ level, will serve as a test of
TC2 with the improvement of measurement precision at the LHC. Then
with all the constraints (including the requirement to explain
$A_{FB}$ at $2\sigma$  level and satisfying the current LHC
measurement of $A_{C}$ at  $2\sigma$ level), we find that the TC2
model can induce sizable triple-top production at the 14 TeV LHC
(the production rate can maximally reach 16 pb). Due to the low SM
backgrounds, the triple-top production can also be a good probe for
TC2 model, complementary to $A_{C}$.
\end{abstract}
\pacs{14.65.Ha,12.60.Nz} \maketitle

\section{INTRODUCTION}
As the heaviest particle observed so far, top quark is speculated
to play an important role in probing new physics beyond the
standard model (SM) \cite{top-review}. Since its discovery, many
of the top quark properties have been firmly established. And most
of the measurements agree well with the SM predictions, except for
the top quark forward-backward asymmetry $A_{FB}$ reported at the
Tevatron \cite{afb-cdf,afb-d0}. The CDF Collaboration measured
$A_{FB}$ based on an integrated luminosity of 5.3 fb$^{-1}$ and
obtained $A^{t}_{FB} = 15.0 \pm 5.5 \%$ \cite{afb-cdf}, which is
larger than the SM prediction 0.056(7)\cite{sm-afb}. Such an
anomaly has been tried to be explained in various
new physics models \cite{afb-review,afb-eft}, such as the
models with $Z'$ \cite{afb-zprime}, $W'$ \cite{afb-wp}
and exotic scalars \cite{shu-scalar,scalar}. In these models, the new
flavor-changing interactions are usually invoked and will lead to
other phenomenologies, such as the like-sign top pair
production \cite{like-sign-top} and single top
production \cite{wu-ac,single-top}, which can be tested
at the LHC.

As a concrete dynamical electroweak symmetry breaking model, the
topcolor-assisted technicolor (TC2) is recently found to be capable
of accounting for the top quark forward-backward asymmetry
\cite{han-tc2}. In TC2 \cite{tc2} the electroweak symmetry breaking
(EWSB) is mainly driven by the technicolor interaction. All ordinary
quark and lepton masses, including a very small portion of the top
quark mass, are provided by the extended technicolor. While the
topcolor interactions give rise to the main part of the top quark
mass and also make small contributions to the EWSB. As is well
known, the topcolor interactions are non-universal \cite{top-color}
and thus cause the tree level flavor-changing neutral-current (FCNC)
interactions for the top quark. These new FCNC interactions between
up quark and top quark can contribute to $A_{FB}$ in the $t\bar{t}$
production at the Tevatron through $t$-channel mediated by the new
scalars (top-pion, top-higgs) and vector bosons (top-rho) at tree
level \cite{han-tc2}. With such a contribution, the discrepancy
between the experimental result and the SM prediction of $A_{FB}$
can be significantly reduced. On the other hand, we note that any
attempts to solve the problem of $A_{FB}$ must satisfy other
experimental measurements on the top quark, such as the recent LHC
measurements on the $t\bar{t}$ total cross section
\cite{lhc-tt-exp}, the differential cross section
\cite{atlas-tt-inv,cms-tt-inv} and the like-sign top pair production
\cite{like-sign-top-cms}. The LHC has also performed a measurement
on the top charge asymmetry \cite{cms-ac,atlas-ac}, which is
considered as a direct test of the anomalous $A_{FB}$ at the LHC
\cite{ac-np}. In our analysis, we will consider all current
constraints from the LHC and Tevatron and examine the correlation
between $A_{FB}$ at the Tevatron and $A_{C}$ at the LHC.

Further more, we note that the FCNC interactions in the TC2 model will
inevitably induce the triple-top ($tt\bar{t}+t\bar{t}\bar{t}$)
production at tree level. Therefore, we will study the TC2 contribution
to the triple-top production at the LHC with $\sqrt{s}=14$ TeV
while requiring TC2 to solve the $A_{FB}$ anomaly.
In the SM, the triple-top can be produced in association with
a $W$ boson or a jet at leading order, and the
production rates are very small \cite{barger-3t}. The pure triple-top
production (without a $W$ or a jet) is forbidden by the GIM mechanism
at tree level, and highly suppressed by the non-diagonal
elements of the Cabibbo-Kobayashi-Maskawa (CKM) matrix at the
one-loop level. On the contrary, in TC2, due to the large couplings
between the top quark and new scalars and vector boson, the pure
triple-top can be copiously produced and may be accessible at
the LHC. Therefore, the triple-top production may provide a new way
to test the TC2 model at the LHC.

This paper is organized as follows. In Sec. II, we briefly outline
the relevant features of the TC2 model. Then in Sec. III we work in
TC2, and present the correlation between $A_{FB}$ and $A_{C}$,
and discuss the triple-top production and its kinematical distributions.
Finally, we draw our conclusion in Sec. IV.

\section{The TC2 model}
Technicolor is a dynamical theory for electroweak symmetry breaking
by condensing fermion bilinears in the vacuum. It can provide a
natural way to explain the weak scale, but is difficult to generate
fermion masses, in particularly, the heavy top quark mass. On the
other hand, the topcolor can produce the large top quark mass but
with an incomplete explanation of EWSB. In order to overcome these
difficulties, the topcolor-assisted technicolor model (TC2)
combining the technicolor interaction with the topcolor interaction
was proposed \cite{tc2}. In this model, there are a number of
pseudo-Goldstone bosons (PGBs) at the weak scale, such as the neutral
top-pion $ \pi_t$. These PGBs can induce the tree level top quark
FCNC interactions arising from the top quark mass term \cite{tc2-3}.
In addition, the new strong interactions can greatly enhance the flavor
conserving top quark interaction with $\pi_t$. Therefore, these new
interactions can not only affect the top pair productions through
$t$ channel at tree level and $s$ channel at loop level, but also
lead to a sizable triple-top production. The relevant interactions
are given by \cite{han-tc2}
\begin{eqnarray}
  \mathcal{L}_{\pi_{t}}  & =i g_{tt\pi_{t}} ( \pi_{t} \bar{t} \gamma^5 t) +i g_{t u \pi_{t}}
  ( \pi_{t} \bar{t}_L u_R) + i g_{t c \pi_{t}}( \pi_{t} \bar{t}_L c_R) +h.c.
\end{eqnarray}
Here $g_{ij\pi_{t}}=m_t/f_{\pi}U^R_{ij}$, $U^R_{ij}$ is the
rotation matrix that transforms the weak eigenstates of the
right-handed up-type quarks to their mass eigenstates, and $f_{\pi}$ is
the vacuum value of top condensate which is about 60 GeV in the NJL
model. However, the indirect constraint from $Z \to b \bar{b}$
requires that $f_{\pi}$ is larger than 100 GeV \cite{tc2-2}. In
addition, the TC2 model also predicts a CP-even scalar called
the top-Higgs ($h_t$). Since the top-Higgs and neutral top-pion are
respectively the real part and imaginary part of one complex scalar
in the linear sigma model, the couplings of the top quark with the
top-Higgs are given by \cite{han-tc2}
\begin{eqnarray}
  \mathcal{L}_{h_t}  &=g_{t t h_t} ( h_t \bar{t} t) + g_{t u h_t} (h_t \bar{t}_L  u_R)
+ g_{t c h_t}( h_t \bar{t}_L c_R) + h.c.
\end{eqnarray}
where $g_{ijh_t}$ is the coupling of $h_t$ to the up-type quarks and
$g_{ijh_t}=g_{ij\pi_{t}}$. In our study, we also consider the
lightest vector excitation of the top condensate, the top-rho,
whose coupling to $t\bar{t}$ is assumed to be a free
parameter \cite{top-rho}. After rotating the up-type quarks to the
mass eigenstates, we can obtain the FCNC interactions between top
quark and top-rho \cite{han-tc2}
\begin{equation}
  \mathcal{L}_\rho   =
g_{tt\rho}\, \rho_\mu \bar{t} \gamma^\mu t
 + g_{t c\rho}\, \rho_\mu \bar{t}_R \gamma^\mu c_R
+ g_{t u\rho}\, \rho_\mu \bar{t}_R \gamma^\mu u_R+h.c.
\end{equation}
where $g_{ij\rho}$ is the coupling of top-rho to the quarks. Due to
the larger masses for the higher excited states of $t\bar{t}$
condensate, we will not discuss them for the purpose of solving the
problem of $A_{FB}$.

As discussed in \cite{han-tc2}, a sizable $u_R-t_R$
mixing does not conflict with the low energy flavor physics
constraints (such as $D-\bar{D}$ mixing and the $B-\bar{B}$ mixing)
given that other flavor mixings are suppressed. In our
analysis, we assume only a large $g_{t u \pi_{t}/h_t/\rho}$ and
$g_{t c \pi_{t}/h_t/\rho}=0$ for simplicity \cite{han-tc2}. Since the
masses of top-Higgs and top-pion can not be calculated from the
theory, we take them as free parameters and further assume they
are equal to avoid the constraints from the like-sign top pair
production at the LHC and Tevatron. For the exited state top-rho,
it is reasonable to set its mass above the ground state
top-Higgs \cite{top-rho}. Although the current data of the LHC and
Tevatron through $WW$ and $ZZ$ channels have excluded a heavy mass
range for the SM Higgs \cite{higgs}, they are not applicable to the
TC2 scalars (top-pion and top-Higgs) because they are responsible
for a small part of the EWSB. For the top-pion, the parity conservation
forbids the couplings of $\pi_{t}WW$ or $\pi_{t}ZZ$; while for the
top-Higss, its couplings to the electroweak gauge bosons $W$ and $Z$
at tree level are suppressed by a factor of $f_{\pi}/v_w$ compared with
the SM Higgs couplings \cite{tc2} ($v_w=174$ GeV is the electroweak vev).

\section{NUMERICAL RESULTS AND DISCUSSIONS}
In our calculations we take the SM parameters as \cite{pdg}
\begin{eqnarray}
m_t=175{\rm ~GeV},~m_{Z}=91.19 {\rm ~GeV},~\sin^{2}\theta_W=0.2228,~\alpha=1/128.
\end{eqnarray}
We use the parton distribution function CTEQ10L \cite{cteq} with
renormalization scale and factorization scale $\mu_R = \mu_F = 2m_t$
for $t\bar{t}$ production and $\mu_R = \mu_F = 3m_t$ for triple-top
production. We scan the parameters in the following ranges
\begin{eqnarray}
&& 200{\rm ~GeV}<m_{\pi_t/h_t}<500 {\rm ~GeV}, ~500{\rm
~GeV}<m_{\rho}<800 {\rm
~GeV}\nonumber \\
&& 1.2<(g_{tt\pi_t},~g_{tt h_t},~g_{tt\rho})<3.3,~~0.5<(g_{t u \pi_t},~g_{tu h_t},~g_{tu\rho})<1.2
\end{eqnarray}
Here the upper bounds on the flavor conserving couplings are based
on the requirement of perturbativity and the lower bounds are taken
from \cite{han-tc2} which are obtained from the consideration of
explaining $A_{FB}$ and avoiding large same sign top production.
In our study, we consider
the following constraints from the LHC and Tevatron:
\begin{itemize}
\item[(i)] The $t\bar{t}$ cross sections:
  \begin{itemize}
 \item Tevatron: based on $4.6 fb^{-1}$ luminosity data,
the $t\bar{t}$ total cross section measured by CDF Collaboration is
$\sigma^{t\bar{t}}_{exp} = 7.50\pm0.31_{stat}\pm
0.34_{syst}\pm0.15_{th}$ pb \cite{tev-tt-exp}. Combining errors in
quadrature, we get $\sigma^{t\bar{t}}_{exp} = 7.50\pm0.48$ pb,
which is in good agreement with the SM prediction $\sigma(t\bar{t})
= 7.5^{+0.5}_{-0.7}$ pb \cite{tev-tt-th};
 \item LHC: recently the CMS Collaboration has reported their combined
results corresponding to an integrated luminosity between $0.8fb^{-1}$ and
$1.1fb^{-1}$, which is $\sigma^{t\bar{t}}_{exp} =
165.8\pm2.2_{stat}\pm10.6_{syst}\pm7.8_{lumi}$ pb \cite{lhc-tt-exp}.
It is consistent with the SM prediction
$\sigma(t\bar{t})=167^{+10+15}_{-17-13}$ pb \cite{lhc-tt-th}.
  \end{itemize}
In our calculations, we require the theoretical prediction
(the SM value plus new physics effects) for the
$t\bar{t}$ cross section to agree with the experimental data
at $2\sigma$ level.

\item[(ii)] The $t\bar{t}$ invariant mass distribution:

Since the new TC2 particles contribute to $t\bar{t}$ production through $t$
channel,
 they may distort $t\bar{t}$ invariant mass distribution.
 \begin{itemize}
 \item Tevatron: we use the data of
 the $t\bar{t}$ invariant mass distribution
 from the CDF Collaboration \cite{tev-tt-inv} and require
 the new physics contribution in each bin to lie within the $2\sigma$ range;
 \item LHC: The high $t\bar{t}$ invariant mass distribution at the LHC has been used
  to exclude a heavy resonance with strong couplings to $t\bar{t}$, such as KK-gluon
  and axigluon \cite{atlas-tt-inv}. In our model, besides through $t$-channel, the top-pion and top-Higgs can
  contribute to the $t\bar{t}$ production through $s$-channel by gluon fusions at loop level.
  We find that they can maximally enhance the differential cross section by about $9\%$,
  which is still within the allowed range of experimental data \cite{cms-tt-inv}.
 \end{itemize}

\item[(iii)] Top+jet resonance in $t\bar{t}$+jets events at the Tevatron:

 We note that the new FCNC interactions will also cause the single top production
$t ({\rm or}~\bar{t})+X$
 with $X$ decaying to $\bar{t} ({\rm or~} t)+jet$.
The CDF Collaboration has recently searched
 for a $t ({\rm or}~\bar{t})+$jet resonance in $t\bar{t}$+jet events
and set an upper limit of $0.61 \sim 0.02$ pb for $m_X=200 \sim 800$ GeV \cite{top+jet}. For our model, when $m_X>2m_t$, the new
 decay mode $X \to t \bar{t}$ will be dominant.
Thus the main constraint on our model is in the mass range $m_X<2m_t$.

\item[(iv)] The like-sign top pair production:

Although the mass degeneracy of top-pion and top-Higgs can partially
escape the like-sign top constraints, the top-rho can also
contribute to $tt$ production.
 \begin{itemize}
  \item Tevatron: the CDF Collaboration has performed an exclusive search for $tt$ production
   and give a upper bound on $tt$ rate:
 $\sigma_{tt}\leq 500$~fb \cite{like-sign-top-cdf};
  \item LHC: Very recently, the CMS Collaboration also published
 their results of $tt$ search for the light $Z'$ model and gave
a upper bound of $0.67$~pb on the $tt$ production rate \cite{like-sign-top-cms}.
 \end{itemize}

\end{itemize}

\subsection{$A_{FB}$ and $A_{C}$ in TC2}
In a given model the prediction for $A^{t}_{FB}$ at the Tevatron
should be correlated to the prediction for $A_{C}(t\bar{t})$ at the LHC.
However, while the Tevatron observed some anomaly for  $A^{t}_{FB}$,
the LHC measurement of  $A_{C}(t\bar{t})$ is in agreement with
the SM prediction \cite{cms-ac,atlas-ac}.
Recently, with an integrated luminosity of 1.09 fb$^{-1}$, the CMS
result is $A^{\rm exp}_{C}(t\bar{t}) =-0.016\pm0.030({\rm
stat.})^{+0.010}_{-0.019}({\rm syst.})$ \cite{cms-ac}, which is
consistent with its SM prediction $A^{\rm
SM}_{C}(t\bar{t})=0.0130(11)$\cite{ac-th}. A similar result is also
reported by the ATLAS Collaboration but with a larger uncertainty
\cite{atlas-ac}. So we use the CMS result in our analysis.

\begin{figure}[htbp]
\includegraphics[width=4in,height=3in]{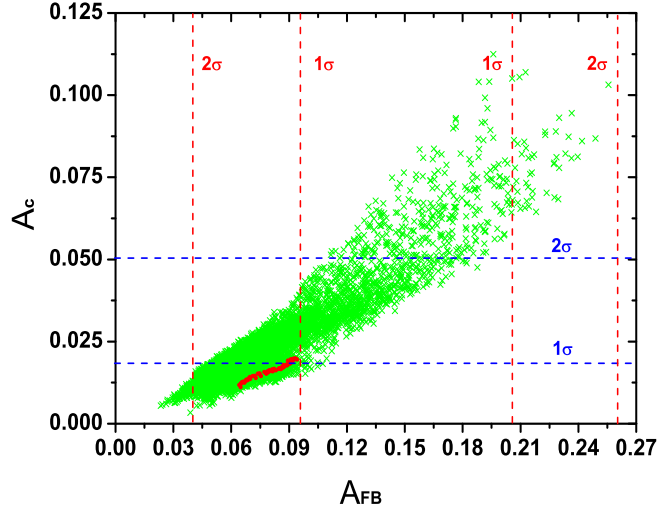}
\vspace{-0.5cm} \caption{Scatter plots of the scanned parameter space
projected on the plane of $A^{t}_{FB}$ (Tevatron) versus $A_{C}$ (LHC):
the dots (red) and the crosses (green) denote respectively the survived
samples with and without the constraints (i-iv).
The horizontal dashed
lines (blue) show the $1\sigma$ and $2\sigma$ upper limits from the
LHC data of $A_{C}$, while the vertical dashed lines (red) show the
$1\sigma$ and 2$\sigma$ regions from the Tevatron data of
$A^{t}_{FB}$. } \label{fig1}
\end{figure}
Firstly we scan over the parameter space of TC2, and then we
calculate $A_{FB}$ and $A_C$ in the parameter space. In Fig.1, we
project parameter space on the plane of $A_{FB}$ versus $A_C$. From
this figure we can see the correlation between $A_{FB}$ and $A_C$.
Generically,  the value of $A_C$ at the LHC is proportional to the
value of $A_{FB}$ at the Tevatron, because the produced top
(anti-top) quark is inclined to go along (against) the valence quark
direction in $t\bar{t}$ production \cite{wu-ac}. In TC2 model there
are two new contributions to $A_{FB}$: one is from the scalars
(top-pion and top-Higgs); the other is from the vector boson
top-rho. Both of them contribute to $t\bar{t}$ production through
$t$-channel, which, due to the Rutherford singularity, can maximally
increase the value of $A_{FB}$ to $13.6\%$. However, only with the
scalars' contribution, $A_{FB}$ can not be enhanced effectively
because of the spin correlation between top and anti-top quarks
\cite{shu-scalar}. Thus, the vector boson top-rho will play an
important role in generating a large $A_{FB}$. We also note that
although the value of $A_{C}$ becomes larger with increasing
$A_{FB}$, most of the samples are still in the $2\sigma$ range of
the experimental value. However, from the red dots which denote the
parameter space survived all constraints (i-iv), we can see most
parameter space has been excluded due to the new like-sign top
experiment results at CMS.

\begin{figure}[htbp]
\includegraphics[width=4in,height=3in]{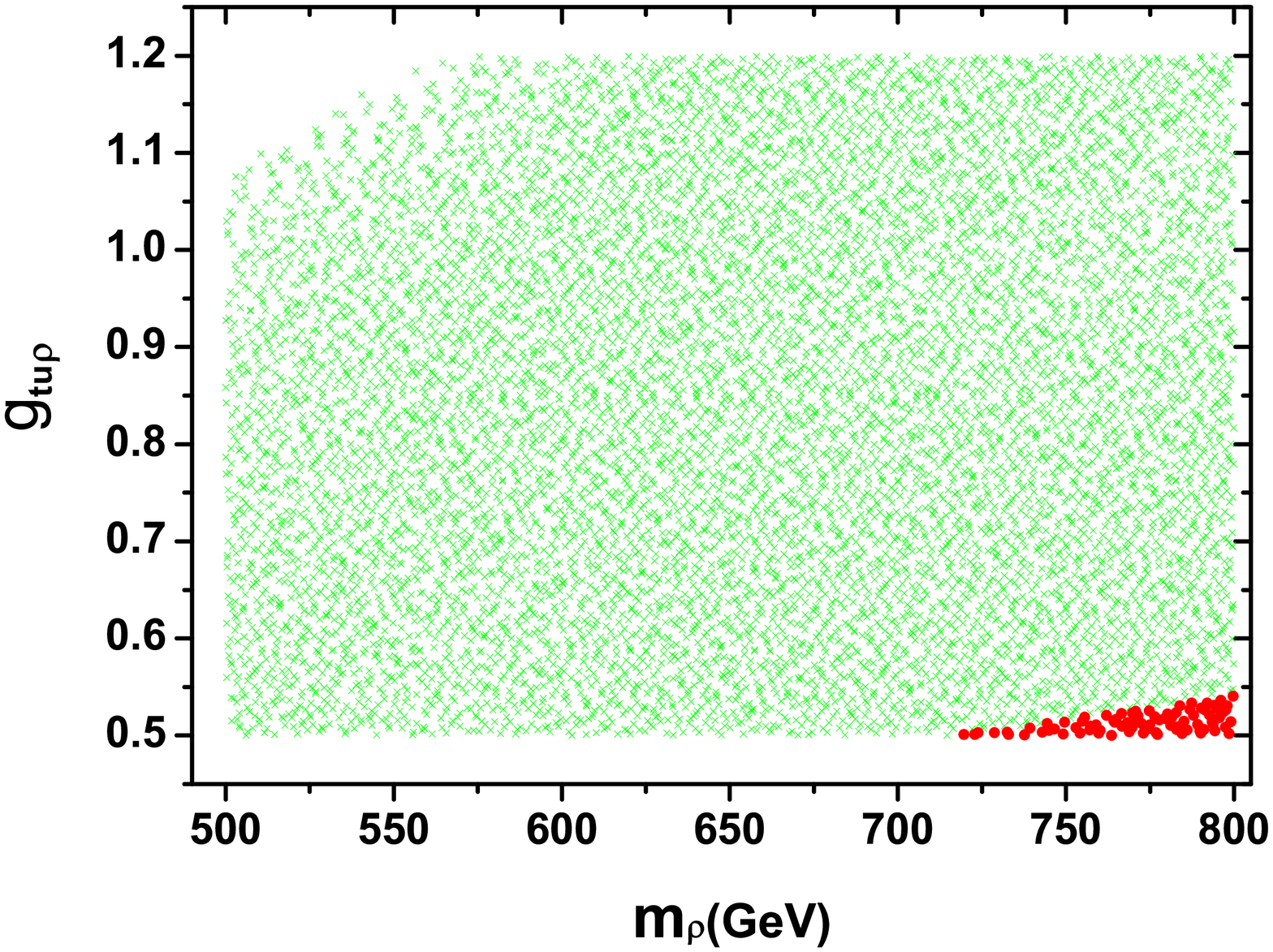}
\vspace{-0.5cm} \caption{The scatter plots of the survived
samples: the crosses (green) are allowed by the Tevatron constraints
while the dots (red) are allowed by all the constraints from
the Tevatron and the LHC.}
\label{fig1}
\end{figure}
In order to show how strong the current LHC constraints are, we in
Fig.2 display two sets of samples: one set (denoted by dots) is
allowed by all constraints from the Tevatron and LHC, and the other
set (denoted by crosses) is allowed by the Tevatron constraints but
not by the LHC constraints. Note that here the Tevatron constraints
include the requirement that the theoretical value of $A_{FB}$
agrees with the experimental data at $2\sigma$ level. We see that
the current LHC constraints are already quite stringent, able to
exclude much of the parameter space allowed by the Tevatron. The
figure shows that the LHC constraints exclude the region with a
large FCNC coupling $g_{tu\rho}$ ($>0.55$) and a light top-rho mass
$m_{\rho}$ ($<730$~GeV). The reason is that a large $g_{tu\rho}$ and
a heavy top-rho mass may lead to a large production rate of
like-sign top pair, which is not allowed by the LHC bound.

\subsection{TC2 contribution to triple-top production at the LHC}
In TC2 model, both the FCNC couplings and the flavor-conserving couplings
are large. This will induce sizable triple-top production at the LHC.
The main contributions are from the $t$-channel diagrams,
as shown in Fig.3.
\begin{figure}[htbp]
\includegraphics[width=4in,height=1.5in]{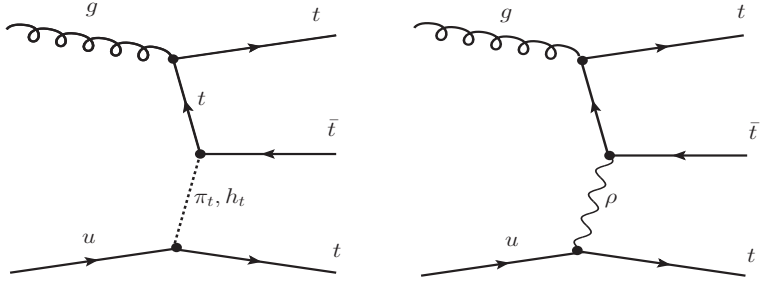}%
\caption{The representative Feynman diagrams for the triple-top
production in TC2 model. } \label{fig3}
\end{figure}

\begin{figure}[htbp]
\includegraphics[width=4in,height=3in]{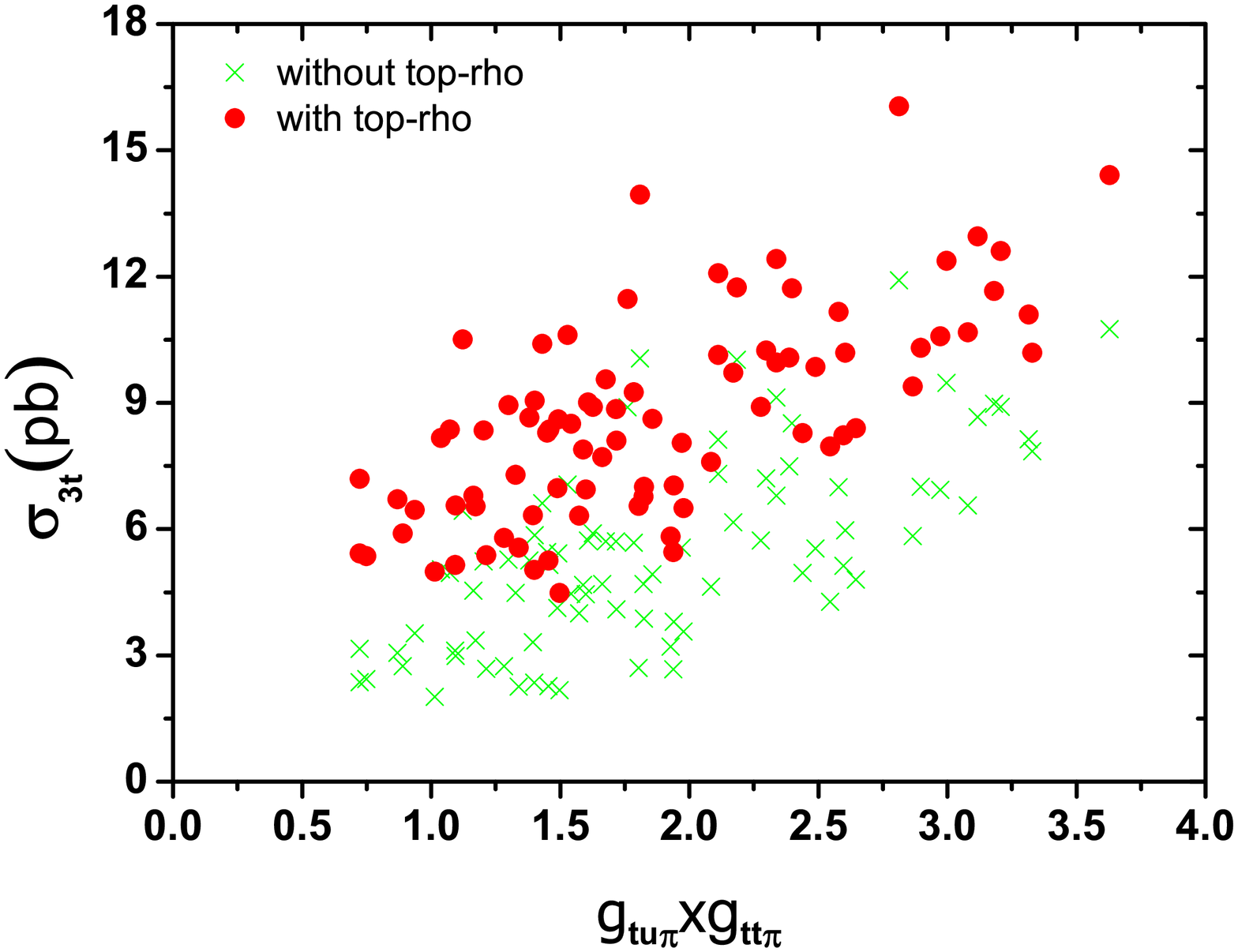}
\vspace*{-0.5cm} \caption{The scatter plots of the TC2 parameter
space survived all the constraints (the Tevatron constraints include
the requirement that the theoretical value of $A_{FB}$ agrees with
the experimental data at $2\sigma$ level, and the LHC constraints
include $A_C$ at $2\sigma$ level), showing the triple-top production
rate at the LHC with $\sqrt{s} = 14$ TeV. The dots (red) shows the
results with the top-rho contribution and the crosses (green)
denotes the cross section without the top-rho contribution.}
\label{fig4}
\end{figure}
Since the triple-top process involves an extra free parameter
$g_{t\bar{t}\rho}$, we generate its values randomly. Other
parameters in the calculation are required to satisfy the
experimental constraints (i-iv) and solve $A_{FB}$ at $2\sigma$
level. The TC2 prediction of the triple-top production rate at the
LHC is shown in Fig.4. From this figure we see that the triple-top
production cross section at the LHC (14 TeV) can maximally reach 12
pb without the top-rho contribution and 16 pb with the top-rho
contribution, which may be detected with a proper reconstruction
technique\cite{barger-3t,multitop,Aguilar}. It should be noted that
since the triple-top production involves an extra coupling
$g_{tt\rho}$ which dose not appear in the $t\bar{t}$ production, the
correlation between the triple-top production rate and $A_{FB}$ or
$A_{C}$ is weak.

In order to provide more information of the triple-top production,
we display some kinematical distributions of final
states by using Madgraph5 \cite{md5}.
For illustration, we take a point in the allowed parameter space
which gives the largest cross section:
\begin{eqnarray} \label{input}
g_{tu\pi}=1.176, g_{tt\pi}=2.39, g_{tu\rho}=0.50,g_{tt\rho}=3.136,
~m_{\pi}=588.12{~\rm GeV}, m_{\rho}=732.87{~\rm GeV}.
\end{eqnarray}
For this set of parameters, some kinematical distributions of the
cross section are shown in Fig.5.
\begin{figure}[htbp]
\includegraphics[width=6.5in,height=2.5in]{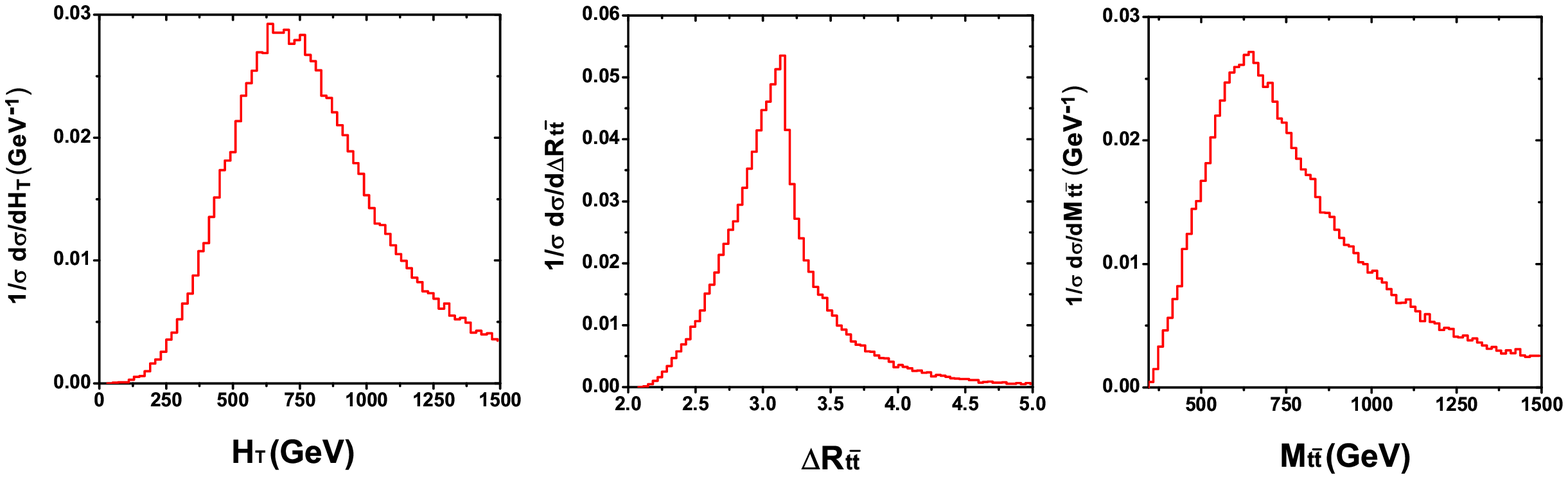}
\vspace*{-1.5cm}
\caption{The $H_T$ (total transverse energy),
$\Delta R_{t\bar{t}}$ (separation between $t$ and $\bar t$)
and $M_{t\bar{t}}$ ($t\bar{t}$ invariant mass) distributions for the
TC2-induced triple-top production at LHC with $\sqrt{s} = 14$ TeV.
The TC2 parameters are fixed in Eq.(\ref{input}).}
\label{fig5}
\end{figure}
From the left panel of Fig.5 we see a peak at about $H_T=500$ GeV
which is higher than the usual SM processes. From the middle panel
we see that the distribution is peaked at a large $\Delta
R_{t\bar{t}}$ near $\pi$, which indicates that the top and anti-top
quarks from the on-shell top-rho tend to go in the opposite
direction. From the right panel we see a peak near the mass of
top-pion or top-higgs($\sim588$ GeV) in the $t\bar{t}$ (they are
from the parent top-pion or top-higgs) invariant mass distribution,
which is caused by the on-shell decay of a top-pion or a top-higgs.
It should be noted that there is not a peak around the mass of
top-rho ($\sim732$ GeV) due to a large decay width of
the top-rho. All these features may be helpful for
detecting the triple-top signal at the LHC.

\section{Conclusion}
In TC2 model we studied its correlated contributions to $A_{FB}$ at
the Tevatron, $A_{C}$ and the triple-top production at the LHC.
Under current constraints on the top quark from the LHC and Tevatron
(such as the total and differential production rates), we scanned
the parameter space of the TC2 model. We found that in the allowed
parameter space the TC2 model can explain the Tevatron measured
$A_{FB}$ at $2\sigma$ level, but meanwhile significantly enhance
$A_{C}$ at the LHC. Such enhanced $A_{C}$, albeit currently allowed
by the LHC measurement at $2\sigma$ level, will serve as a test of
TC2 with the improvement of measurement precision at the LHC. Then
with all the constraints (including the requirement to explain
$A_{FB}$ at $2\sigma$  level and satisfying the current LHC
measurement of $A_{C}$ at  $2\sigma$ level), we found that TC2 model
can induce sizable triple-top production at the 14 TeV LHC (the
production rate can maximally reach 16 pb). Due to the low SM
backgrounds, the triple-top production can also be a good probe for
TC2 model, complementary to  $A_{C}$.
\section{Note added}
After we finished our manuscript, the CMS Collaboration
published a search for events with three or more isolated leptons in
pp collisions at $\sqrt{s}$ = 7 TeV with an integrated luminosity of
4.98 $fb^{-1}$\cite{multilepton}. Since our triple-top production can also
give a final state with three leptons, this search may be relevant to our
study. So we calculated the triple-top production for $\sqrt{s}$ = 7 TeV
(8 TeV) and found that the production rate can maximally reach 2 pb (3 pb),
which, without any cut, can give the tri-lepton events below 200 (300).
We checked that such a number of events is allowed by the CMS results.

\section*{Acknowledgement}
Chengcheng Han was supported by a visitor program of Henan Normal University, during which
this work was finished. This work was supported in part by the National Natural Science Foundation
of China (NNSFC) under grant Nos. 10821504 and 11135003, by the Project of Knowledge Innovation
Program (PKIP) of Chinese Academy of Sciences under grant No. KJCX2.YW.W10. and by the Startup
Foundation for Doctors of Henan Normal University under contract No.11112.

\end{document}